# THE PURPOSE OF ASTRONOMY

Emmanuel Davoust

URA 285, Observatoire Midi-Pyrénées, 14 Avenue Belin, 31400 Toulouse, France

ABSTRACT

This is a presentation of the purpose of astronomy in the context of modern society. After exposing two misconceptions about astronomy, I detail its role in five domains, certified knowledge, incorporated abilities, innovations, collective goods, and popular science; with each domain is associated an institution, an incentive, and a method of evaluation. Finally, I point out the role of astronomy as a source of inspiration in other fields than science.

## 1. A poorly phrased question

"What is the purpose of astronomy?" It seems to be a very simple question. But the answer is difficult to find. This is partly because the question is based on out-of-date ideas about the real meaning of astronomy in particular, and fundamental research in general.

If I had been asked the question one or two centuries ago, at the time that astronomy was an applied science, I would have had no difficulty answering the question in the terms in which it had been posed.

I would first of all have mentioned geography, and would have recalled the long journeys – of Charles de La Condamine, of Louis-Antoine de Bougainville, – to precisely measure the longitudes and latitudes of various points of the globe. I would have mentioned maritime navigation, sailors using sextants and position tables of the moon and bright stars in order to establish their position at sea. I would have mentioned metrology and the adventures of astronomers Jean-Baptiste Delambre and Pierre Méchain as they travelled all over France during the Revolution in order to establish the length of the meridian, and, consequently, that of the meter, which is one-forty millionth of the meridian. I would have recalled that the Observatory in Besançon, near Switzerland, where I started my career, was founded over a century ago to check the accuracy of the local watchmakers' products.

But this period is over. Positioning, maritime and aerial navigation are now carried out with beacons and artificial satellites. The meter's definition has changed and is now based on the wavelength of the transition of a krypton atom. The definition of the second of time has changed too; it is given by atomic clocks running on cesium atoms.

Does this mean that astronomy no longer serves any purpose? That it has, at most, a cultural role – that of teaching us what a black hole is, or how the Universe formed and how it will end? Surely that in itself does not justify governments paying the salaries of armies of astronomers permanently watching the sky with large numbers of ever more expensive telescopes? It is even less justifiable in a period of crisis and unemployment, at a time when the budgets for health, transportation, education are in deficit.

Or do governments continue funding fundamental research in the hope that one day, in fifty, a hundred or a thousand years, it will finally produce something that can be sold, like computers, or something useful, like electricity? But remember that politicians usually want their investments to give returns on a short term basis. Even if fundamental research sometimes has "useful" results, nobody has ever demonstrated that it is the cheapest and most direct way to technological progress, and from that point of view, a country like Japan prefers to invest massively in applied research.

It is therefore obvious that the question is poorly phrased. In order to answer, I will first expose some misconceptions (Callon, 1995): that of astronomy as an isolated science with well-defined frontiers, that can be independently evaluated, then that of astronomy as an abstract science, that may or may not be applicable to the country's other activities.

I will then pose the question in slightly different terms, placing astronomy, and more generally, fundamental research, in a modern context, one where they are not isolated from other activities that make us live. Fundamental research is indeed an integral part of social and economic activity, which forms a dense network of relations in which circulate and act not only concepts and ideas, but also people, instruments and institutions.

## 2. A first misconception: the territoriality of research

The first difficulty that arises when we want to evaluate astronomy is that it cannot be isolated from other sciences. Of course, astronomy can be defined quite precisely, like the science of everything that is outside Earth. However, in order to investigate this subject, astronomers need other sciences, especially physics, but also computer science, chemistry, mathematics, biology, etc.

Conversely, scientists in other disciplines take advantage of the fact that the Universe is an exceptional laboratory, containing physical extremes, of temperature, pressure and density, in order to study matter in situations that are impossible to duplicate in a lab, such as the extremely cold and almost empty state in molecular clouds, or the infinite pressure and density in neutron stars.

There are therefore a number of scientists who work in areas bordering on astronomy, and who do not really know whether they are astronomers, chemists or nuclear physicists. Because of its complexity, fundamental research has become multidisciplinary.

Let me give a few concrete examples.

One of my colleagues is investigating why a particular atom, the calcium of atomic number 48, exists more abundantly in the Universe than in its isotope, calcium 46. As part of her research, she carries out experiments on a particle accelerator. The reactions found in this laboratory will enable her to explain what goes on in the cosmic laboratory.

Another colleague is interested in the physics of the interstellar medium. In this medium, there are molecules that can be purchased in a bottle at a drugstore, but also many other molecules that laboratory chemists have never seen. To check the identity of molecules discovered in the interstellar medium, calculations and experiments based on pure chemistry have to be made, with tools and concepts that were developed by chemists. This work is done in collaboration with chemists who are interested in the same problems for other reasons.

The next Cassini mission to Saturn will include the launching of a small probe, Huyghens, to explore the atmosphere of the satellite Titan. In this atmosphere, scientists hope to find prebiotic molecules, and perhaps even biological compounds. Preparing this mission requires the collaboration of specialists of all kinds: plasma physicists, biologists, computer scientists, specialists in prebiotic chemistry, planetary atmospheres, telemetry, etc. So, paradoxically, the hyperspecialization of research in fact brings about a split in classical disciplines and the appearance of interfaces, of fruitful exchanges among neighboring disciplines, creating new ones.

This complexity in research, this overlap of disciplines, this mixing of tasks, reflect a basic characteristic of modern research – its organization into networks. The idea of a network is essential to the understanding of the nature of research activity. We will return to this point later.

We can add that the formal division of science that currently exists is a practical and administrative necessity, but which no longer corresponds to reality. The managers of fundamental research want this division in order to administrate scientists and their activity, for their statistics, to be able to say that there are so many astronomers, so many nuclear physicists, so many chemists; that each group produces so many articles and theses, has so many students, has a certain budget, etc. The researchers themselves must belong to a laboratory.

## 3. A second misconception, the linear model of research

In an incomplete picture of fundamental sciences, astronomy is a set of laws, empirical rules, postulates, a set of abstract concepts, that engineers, industrialists, doctors or other actors of economic life can pick off the shelf and use to solve a concrete problem, to build a machine useful to man, to increase the productivity of a plantation, to check an epidemic, to prevent a natural catastrophe, etc.

We would thus have the following model of research:

$$\text{Astronomy} \longrightarrow \text{Concepts} \longrightarrow \text{``useful'' applications}$$

From this point of view, the only use of fundamental research is to increase the collection of knowledge which, sooner or later, will find a "useful" application, thanks to someone ingenious and curious.

This point of view is the reflection of what the history of astronomy and the history of science have been teaching us for a long time. They have always been histories of concepts and ideas. But this history misinterprets three-quarters of what constitutes science. Even if concepts have an important role to play, men, instruments and institutions, as well as the links between them and the networks that they form, are just as indispensible.

It might seem obvious that concepts did not create themselves, without scientists nor instruments, so why erase them from the picture, as though they were simply vehicles or accessories to knowledge? Idealism or nearsightedness? Because, most of the time, they are unavoidable. Without telescopes, Galileo and all the astronomers who came after him would probably not have discovered anything. And if the astronomers of past generations had not transmitted their abilities as well as their knowledge, the latter would have served no purpose.

In astronomy, as in other sciences, concepts are useless if they are separated from competence, ability, instruments, and the relational networks, that gave them birth. Just as a mushroom cannot grow without an underground network of mycelium, so concepts cannot produce "useful" applications if they are taken out of context. Therefore, in order to evaluate the role of astronomy in society, we have to reconstitute the context in which the knowledge was produced, and evaluate one by one, according to their specific criteria, the different actors in the research process.

So how then can we evaluate a discipline if we cannot isolate it from the context of fundamental research, and if we have to take into account all the actors – living, inanimate, or abstract – who play a role? Clearly, the task is difficult, even after misunderstandings have been cleared up. To clarify things, I now suggest a new division of the picture, probably also reductionist, but very handy when we try to evaluate the role of fundamental research, and in particular of astronomy, a division that takes into account the multiple dimensions of research and its network organization.

## 4. A compass card for research

The compass card for research was suggested by sociologists at the Ecole Nationale Supérieure des Mines in Paris, as a description of fundamental research, taking all its parameters into account (Callon et al., 1994). Let us apply this diagram to astronomy, in order to evaluate its role in the different fields of our society – social, cultural, economic, political, and even artistic.

Each direction on the compass card corresponds to an institution, a kind of incentive, and a method of evaluation. The prestige of astronomy and astronomers' careers are not evaluated in the same way. In one case, we count the votes given to politicians who suppported a scientific project; in the other case, we consider the number, the quality and impact of a scientist's publications. The same motives do not lead to the discovery of new galaxies, to a doctoral thesis or to technological innovations. The scientist is motivated by the thirst for knowledge, the student wants to obtain a degree, the industrialist wants to earn money.

*4.1 Certified knowledge*

The production of new knowledge is certainly astronomy's first objective. We try to explain a star's birth, evolution and death, predict the future of the Universe, describe the shape of the Milky Way without being able to examine it from the outside – in other words, make the Universe understandable.

But new knowledge is not immediately accepted by the whole scientific community as soon as it is announced. Before the validity of any knowledge is accepted, its authors must follow a strict protocol. They must make a detailed technical report, following strict rules about working hypotheses, presenting results, placing the work within current research, whose antecedents must be cited. The report is submitted for publication in a specialized journal that consults experts in the given field (the referees), who give their opinion about the quality of the work. Once published, the work is carried on by others, who confirm (or sometimes invalidate) the result, extend it to other categories of phenomena, etc. Work is thus never totally accepted; it can always be called into question – falsified, as Karl Popper would say.

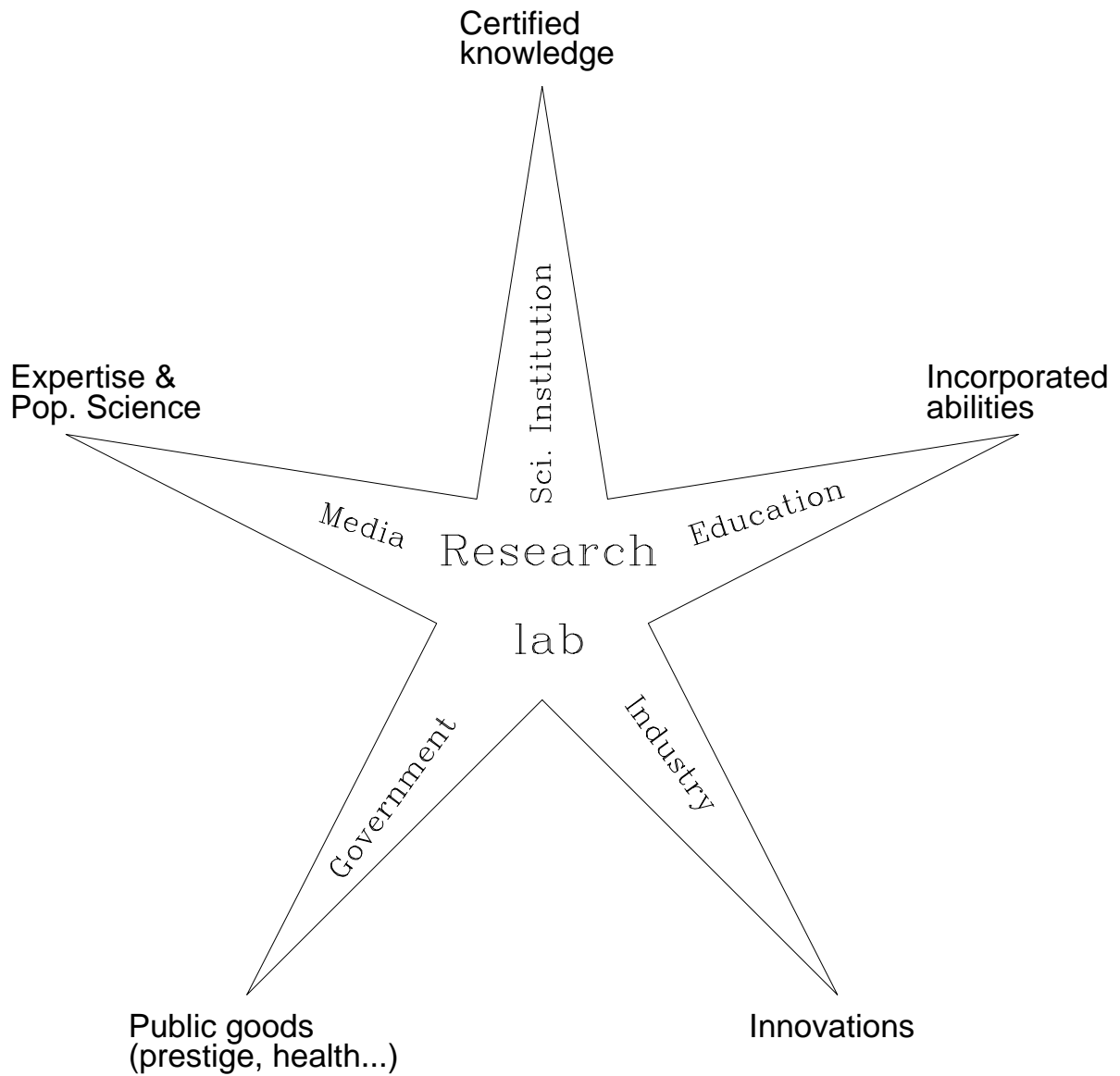

Fig. 1 : The compass card for fundamental research

The important thing for a scientist is to lay claim to a discovery. The principal moving force behind research of new knowledge is peer recognition. This recognition affects the scientist's reputation, his career; he may be honored, given a professorship in a university or the direction of a laboratory.

This need to claim a discovery sometimes leads to distortion or deviation from the classical validation procedure. Since generally six months or a year can pass between the submission of an article and its publication, and that during that time leaks may occur, some scientists, convinced that their discovery is sensational, short-circuit the usual procedure by publishing a press release, that is then taken up by the non-specialist press. Thus, we sometimes read in the daily paper, or hear on the radio or television news, that a certain group discovered a new planet, or a quasar on the frontiers of the Universe, without the discovery's going through the usual filtering process.

Managing the resources of our discipline requires a more global evaluation, that of a scientific theme, or a country. In such a case, bibliometric evaluation is more and more often used (Davoust and Schmadel, 1987). Indeed, because of their number and their impact, publications reveal the success of different research fields. In this way, the frontiers of research can be mapped, networks of collaborators can be emphasized, by counting how many times certain articles (in a specific field, or of a specific country, etc.) are cited by other articles. This is one of the applications of a new science – scientometrics – which measures science.

The study of the history of science, which shows the influence on the evolution of astronomy of a person like Antoine Le Verrier or Arthur Eddington, of a project like *la Carte du Ciel*, of an instrument like the Palomar 200-inch telescope, can also contribute to the evaluation of our discipline, and to learn from the past in order to prepare the future. There is no doubt that the sky survey of *la Carte du Ciel*, which kept the Observatories of Paris, Algiers, Bordeaux and Toulouse busy for decades, was an overly ambitious project for the human and technical means of the time. It is one of the main reasons why the beginnings of astrophysics and extragalactic astronomy did not include the French.

*4.2 Incorporated abilities*

All the astronomers' knowledge serves no purpose if it is taken out of its environment. You can parachute an astronomy textbook into the middle of Siberia, but nothing will happen if you do not at the same time drop astronomers, telescopes, computers, and their whole network of relationships with the outside world.

The production of abilities, generally incorporated in students, trainees at different levels, is an important byproduct of astronomy. Abilities are also incorporated in machines, which are programmed for certain tasks. These abilities will not necessarily be used for astronomical research, because the students may work for private companies and the image treatment software can be used by doctors or geographers. The institution that corresponds to this dimension of research is the teaching establishment – university, engineering school, etc. – and the incentive is generally the diploma received at the end of studies or the training course. I emphasize the fact that these abilities do not necessarily benefit astronomy; training *through* research is one of our missions.

*4.3 Innovations*

Astronomy contributes indirectly to industrial innovation because astronomers tend to ask industrialists to build them instruments at the technical limits, and even often beyond the limits of what is possible, of sensitivity, reliability, precision, and size. Our vocation is to measure extremely weak radiation sources, so we need ultrasensitive instruments with the least possible background noise. Sometimes there are constraints on heat, size or weight. In short, we often ask industrialists to make us a world's wonder.

Even if the instruments supplied to us can rarely be used in other fields, they force industrialists to change their habits, to develop new methods which may be useful elsewhere. Simply in economic terms, the orders for very expensive instruments stimulate industry; through public research, the State acts as a kind of patron of industry.

Some technology developed for astronomy can be applied to other fields. CCD technology, electronic image receivers, and computerized image treatment can be applied to satellite or medical imagery. In 1987-88, two medical scientists studied the eye's retina at the astronomical Observatory in Toulouse, on images taken with astronomical CCD's, and with software designed for astronomical images.

*4.4 Collective goods*

Politicians are directly interested in astronomy because it is something that can bring prestige to the country. Astronomy is part of big science. Ground-based and space-borne observing instruments are very expensive and the technical feat involved in putting them into operation and getting extraordinary images of the surface of Mars or Neptune, for example, will have a positive effect on the prestige of our science, as well as on the country's other activities.

This is why the Western countries agree to finance huge telescopes, like Chile's *Very Large Telescope*, or else the four (and soon five) antenna interferometer at the Institut de Radio-Astronomie Millimetrique on the Bures plateau in the Alps.

But these prestigious acts can backfire on their authors when the project fails (like the Mars Observer satellite in 1993), or if there is a serious defect (as in the Hubble Space Telescope in 1990). There are also limits to what a country can finance; the United States Congress just eliminated the credits needed to continue building the SSC, the super collider in Texas, after two of the ten billion dollars of the projected cost were already spent.

If there are no repercussions in prestige, the government usually makes sure that the big projects that it finances will have consequences for industrialists' order books. Thus the VLT's 8-meter mirrors were ordered from REOSC, a company in the Paris area. In fact, obtaining manufacturing contracts for national industries is a *sine qua non* condition for a country's joining a large international project.

*4.5 Expertise*

Astronomers' expertise is often required for anything dealing with calendars. Insurance companies need to know whether it was still daylight on such a day at such a time, when there was an accident. I even received a phone call from a lady who wanted to know if it would be dark on a certain day at 6 p.m., so that she would know whether a flash would be needed to take photographs at her daughter's wedding! Some religious events are linked to the lunar calendar, such as Easter or Ramadan.

I mentioned above that astronomy no longer measures seconds most precisely. But for historical reasons, astronomers have maintained this expertise. They still manage the measuring of the second at the International Hour Office (BIH). Paris Observatory still gives the public the exact time, and French observatories still participate in French atomic time.

Astronomers also help to correct calendars. The slightly erratic motions of Earth's axis of rotation about an average position shorten or lengthen the duration of its rotation, and the BIH sometimes decides to add or subtract a second from the legal hour in order to readjust it to solar time.

*4.6 Popularizing Science*

Popularization is certainly astronomy's most obvious spin-off. The public is greatly interested in the nature and history of the Universe. The role of astronomers would be incomplete if they only spread their knowledge among their colleagues. Since taxpayers finance this research, they have the right to profit from the acquired knowledge, and to know how public money destined for research has been invested.

Popularization can also be distorted when it becomes a "showcase" of science instead of a sharing of information, when scientists only show beautiful images and unexplained illegible formulae, or impressive instruments without explaining how they work, with the excuse that the public would not understand anyway.

Without leaving the field of science, we can talk about beautiful theories, a property that scientists often take into account when evaluating their theories. Unfortunately, this beauty is often only accessible to specialists. On the other hand, anybody can appreciate the beauty of pictures of the sky, or the majesty of a spiral galaxy unwinding its arms in the cosmic void. This is probably the aspect of research that is most easily shared by the public.

## 5. Astronomy – the source of artistic, literary or religious creativity

I would add a sixth aspect to the five preceding ones: the role of astronomy as inspiration in fields that have nothing to do with science. I mean artistic, literary, or religious creativity.

Astronomy is an inexhaustible source of creativity in all the fields of the arts, especially in science-fiction literature, poetry, painting and plastic arts. The sky has always been the location of the sublime, or transcendence, and this is why so many works of art have been inspired by it.

Show pictures of the solar system to a class in primary school, suggest that they write a poem or make a drawing inspired by what they saw, and they will do wonders.

A composer from Toulouse, Jean Girves, was inspired by sounds from a microphone linked to a radiotelescope observing pulsars, stars emitting radio waves with a period of one second or so, to make an original musical creation. The composer's art was to create a musical sound based on rhythms that are not a harmonious whole, because the periods of the selected pulsars are not commensurable.

These three examples are better than a long dissertation on the subject. I would simply like to add that there is no question of making judgements about their scientific value. They usually have none, and they should only be appreciated for what they are – works of art.

Astronomy can also be a source of religious inspiration, if knowledge of the mechanisms of cosmic clockwork incite us to look for the Watchmaker, or simply to meditate upon the beauty of creation. But we should not look for collaboration between astronomy and religion, for example to interpret the Bible. This attitude, concordism, is considered erroneous by the Church. Both astronomy and religion offer an explanation of the Universe, but in spheres that are different – complementary and not conflictual, at least in my point of view. In one case, it is a rational explanation; in the other, a dogma, a revelation, another perception of reality.